\documentclass[12pt]{article}
\usepackage{amsmath,amssymb,theorem,cite,epsfig,url,psfrag,eepic,amsmath,mathtools,mathrsfs,amsbsy,dsfont}
\usepackage{indentfirst}
\usepackage{graphicx}
\usepackage{tikz,pgfplots}
\usepackage{bm,upgreek}
\usepackage[bookmarks=true,hyperfigures=true,colorlinks=true,linkcolor=black,citecolor=black,urlcolor=black,bookmarksnumbered,hidelinks]{hyperref}
\advance \topmargin by -\headheight
\advance \topmargin by -\headsep     
\evensidemargin \oddsidemargin
\def\Maketitle{{\def\newpage{}\maketitle}}
\begin{document}
\rightline{\texttt{\today}}
\title{\textbf{Bethe ansatz equations\\ for quantum $\mathcal{N}=2$ KdV systems}\vspace*{.3cm}}
\date{}
\author{Dmitry Kolyaskin$^{1}$ and Alexey Litvinov$^{2,3}$\\[\medskipamount]
\parbox[t]{0.85\textwidth}{\normalsize\it\centerline{1. Australian National University, Canberra ACT 2601, Australia}}\\
\parbox[t]{0.85\textwidth}{\normalsize\it\centerline{2. Krichever Center for Advanced Studies, Skolkovo Institute of Science and Technology, 143026 Moscow, Russia}}
\\
\parbox[t]{0.85\textwidth}{\normalsize\it\centerline{3. Landau Institute for Theoretical Physics, 142432 Chernogolovka, Russia}}}
\Maketitle
\begin{abstract}
Based on our previous studies of affine Yangian of $\widehat{\mathfrak{gl}}(1|1)$ we propose Bethe ansatz equations for the spectrum of $\mathcal{N}=2$ quantum KdV systems.
\end{abstract}
\section{Introduction}
The systematic study of integrable models derived from corresponding conformal field theories (CFTs) was initiated by A. B. Zamolodchikov in \cite{Zamolodchikov:1989zs}. It was shown that for particular perturbing operators, an infinite number of mutually commuting integrals of motion (IMs) are preserved. This phenomenon can be illustrated by a large class of 2D CFTs, which are described by Toda action
\begin{eqnarray} \label{Toda action}
S_{0}=\int \bigg(\dfrac{1}{8\pi}(\partial_{\mu} \boldsymbol{\varphi}, \partial_{\mu} \boldsymbol{\varphi})+\Lambda\sum\limits_{r=1}^{N}e^{b(\boldsymbol{\alpha}_r, \boldsymbol{\varphi})}\bigg)d^{2}x,
\end{eqnarray}
where $\boldsymbol{\varphi}=(\varphi_{1},...,\varphi_{N})$ is the $N$-component bosonic field and $\{\boldsymbol{\alpha}_1,..., \boldsymbol{\alpha}_N\}$ is the set of linearly independent vectors. It is known that under certain conditions on these vectors the model contains a larger symmetry. Namely, it turns out that this theory has an extended conformal symmetry, which is also called $W$ - algebra \cite{Zamolodchikov:1985wn}, provided that $\boldsymbol{\alpha}_{r}$ are the simple roots of a semi-simple Lie algebra $\mathfrak{g}$ of rank $N$. Corresponding integrable QFT is described by the perturbation
\begin{eqnarray}
S_{0}\rightarrow S=S_{0}+\lambda\int e^{b(\boldsymbol{\alpha}_0, \boldsymbol{\varphi})}d^{2}x
\end{eqnarray}
where the additional vector $\boldsymbol{\alpha}_0$ is either untwisted  or twisted affine root of the corresponding affine system. Integrability implies that the resulting theory survives an infinite set of mutually commuting Integrals of Motion
\begin{eqnarray}
\boldsymbol{I}_m(\lambda)=\boldsymbol{I}_m+O(\lambda), \quad \bar{\boldsymbol{I}}_m(\lambda)=\bar{\boldsymbol{I}}_m+O(\lambda)
\end{eqnarray}
perturbative in $\lambda$, where $\boldsymbol{I}_m, \bar{\boldsymbol{I}}_m$ are the Integrals of Motion related to $W-$algebra in CFT. 

Another example of integrable perturbation from CFT and the main object of the current work is the supersymmetric generalization of Toda system whose action is of the form \begin{equation}\label{SUSY Toda action}
S_{0}^{SUSY}=\dfrac{1}{2\pi}\int \bigg((D\mathbf{\Phi}, \bar{D}\mathbf{\Phi})+\dfrac{2\pi\Lambda}{b^{2}}\sum\limits_{r=1}^{N}e^{b(\boldsymbol{\alpha}_{r}, \mathbf{\Phi})} \bigg)d^{2}zd^{2}\theta,
\end{equation}
where $\boldsymbol{\Phi}=(\Phi_{1},\dots,\Phi_{N})$ is the bosonic $N-$component superfield 
\begin{equation}
\boldsymbol{\Phi}(z,\bar{z},\theta,\bar{\theta})=\boldsymbol{\varphi}(z,\bar{z})+\theta\boldsymbol{\psi}+\bar{\theta}\bar{\boldsymbol{\psi}}+\theta\bar{\theta}\boldsymbol{F}
\end{equation}
and superderivatives are defined as
\begin{equation}
D=\frac{\partial}{\partial\theta}-\theta\partial,\qquad
\bar{D}=\frac{\partial}{\partial\bar{\theta}}-\bar{\theta}\bar{\partial}.
\end{equation}
In this case, integrability implies that the set of vectors $\{\boldsymbol{\alpha}_{r}\}$ is the completely odd system of simple roots of some Lie superalgebra (which admits such a set). This model is both classical and quantum integrable ultimately invariant under $\mathcal{N}=1$ supersymmetric transformations  \cite{Evans:1990qq, Komata:1990cb}. In particular, in this work we focus on the theories based on $A(n|n)$ Lie superalgebras (for which one can choose a purely odd system of simple roots). Actually, in addition to explicit $N=1$ supersymmetry, $A(n|n)$ Toda theory is $\mathcal{N}=2$ supersymmetric \cite{Komata:1990cb}. The integrable perturbation of the action (\ref{SUSY Toda action}) is realized by the affine untwisted extension $\hat{A}(n|n)$
\begin{eqnarray}
S_{0}^{SUSY}\rightarrow S^{SUSY}=S_{0}^{SUSY}+\lambda\int e^{b(\boldsymbol{\alpha_{0}}, \boldsymbol{\Phi})}d^{2}x,
\end{eqnarray}
where $\boldsymbol{\alpha}_0$ is the lowest fermionic root and the perturbed theory is still $N=2$ supersymmetric \cite{Evans:1990qq}.

A separate problem of the integrable quantum field theories arises regarding simultaneous diagonalization of corresponding IMs. The systematic study of this question was initiated in the series of work by Bazhanov, Lukyanov and Zamolodchikov \cite{Bazhanov:1994ft, Bazhanov:1996dr, Bazhanov:1998dq}. Their research was devoted to the basic example of the integrable model in CFT - the quantum KdV system, which corresponds to $\mathfrak{sl}_2$ case. Among the main results of the study was the construction of the so-called $\boldsymbol{T}$-operator, which plays the role of the generating function for the local and nonlocal IMs, and obtaining functional relations on it. And finally, the work \cite{Bazhanov:2004fk}  of the same authors drew a line in the study of the spectrum of IMs for $\mathfrak{sl}_2$ model. Their approach was based on the connection between integrable systems of CFT and spectral theory of particular differential equation \cite{Dorey:1998pt,Dorey:1999uk,Bazhanov:1998wj}, and the eigenvalues of IMs were expressed through the solutions to a particular system of algebraic equations.
 
There are also different approaches to the diagonalization problem including those based on the affine Yangian Symmetry \cite{Maulik:2012wi}, which, in particular, we apply to finding the spectra of $\mathcal{N}=2$ KdV system in the current work. This approach allows to obtain Bethe ansatz equations, and it is expected that the corresponding eigenvalues of local and quasi-local IMs are expected to be symmetric polynomials in Bethe roots (see also \cite{Litvinov:2013zda,Alfimov:2014qua}). So, for example, based on Yangian symmetry, Bethe ansatz equations were derived, and diagonalization problem was solved for the $Y(\widehat{\mathfrak{gl}}(1))$ case \cite{Litvinov:2020zeq} and for $Y(\widehat{\mathfrak{gl}}(2))$ case \cite{Chistyakova:2021yyd}. Following similar approach, we find the spectra for local and quasi-local IMs of quantum $\mathcal{N}=2$ KdV hierarchy. Our present note will heavily draws on our recent paper \cite{Kolyaskin:2022tqi}, where the current realization of affine Yangian $Y(\widehat{\mathfrak{gl}}(1|1))$ was obtained, and can be considered as sequel. 

The content of the paper is organized as follows. In section \ref{Quantum KdV} we introduce quantum $\mathcal{N}=2$ KdV systems, as they appear in field theory. In section \ref{BAE-section} we formulate the spectral problem for the simplest model with $\mathcal{N}=2$ Virasoro symmetry and provide Bethe ansatz equations for its spectrum. In section \ref{Ygl(1|1)-ansatz} we sketch the relation between $\mathcal{N}=2$ KdV systems and $\textrm{Y}(\widehat{\mathfrak{gl}}(1|1))$ introduced in \cite{Kolyaskin:2022tqi}. We also sketch the derivation of Bethe ansatz equations developed in \cite{Litvinov:2020zeq,Chistyakova:2021yyd}.
\section{Quantum $\mathcal{N}=2$ KdV systems}\label{Quantum KdV}
The most natural way to treat the affine supersymmetric $\mathfrak{sl}(n|n)$ Toda field theory ($A(n-1|n-1)$ Toda) is to consider it as an integrable perturbation of the conformal $\mathfrak{sl}(n|n-1)$ Toda field theory by two exponential operators corresponding to auxiliary $\mathfrak{sl}(n|n)$ root and to affine root of $\mathfrak{sl}(n|n)$.  It is well known that $\mathfrak{sl}(n|n-1)$ Toda field theory is invariant under  the action of $\mathcal{N}=2$ $W_{n}$-algebra \cite{Ito:1991wb}. This class of algebras contain one $\mathcal{N}=2$ multiplet of holomorphic currents for each integers spin $s=1,\dots,n-1$. Perturbing $\mathfrak{sl}(n|n-1)$ by two exponential operators  breaks down the $W_n$ symmetry, but survives an infinite tower of Integrals of Motion. The ultraviolet limit of these Integrals of Motion splits into two sets holomorphic and antiholomorphic conformal IM's. Each of these sets corresponds to integrable system that we will call quantum $\mathcal{N}=2$ $\text{KdV}_n$ system with $n\geq2$. In the following we will consider in details the case of $n=2$.  

Consider $N=2$ Virasoro algebra, which is generated by the currents of spins $(1,\frac{3}{2},\frac{3}{2},2)$
\begin{equation}\label{N=2-Virasoro-currents}
    J(x)=\sum_{n\in\mathbb{Z}}J_ne^{-inx},\quad G^{\pm}(x)=\sum_{r}G^{\pm}_re^{-irx}
    \quad\text{and}\quad T(x)=\sum_{n\in\mathbb{Z}}L_ne^{-inx}-\frac{c}{24},
\end{equation}
with commutation relations
\begin{equation}\label{N=2-Virasoro}
    \begin{aligned}
       &[L_m,L_n]=(m-n)L_{m+n}+\frac{c}{12}(m^3-m)\delta_{m,-n},\\
       &[J_m,J_n]=\frac{c}{3}m\delta_{m,-n},\\
       &[L_m,J_n]=-nJ_{m+n},\\
       &\{G^{+}_r,G^{-}_s\}=L_{r+s}+\frac{r-s}{2}J_{r+s}+\frac{c}{6}\left(r^2-\frac{1}{4}\right)\delta_{r,-s},\\
       &[L_m,G^{\pm}_r]=\left(\frac{m}{2}-r\right)G^{\pm}_{r+m},\\
       &[J_m,G^{\pm}_r]=\pm G^{\pm}_{r+m}.
    \end{aligned}
\end{equation}
This algebra acts in the Hilbert space of some $N=2$ superconformal field theory which splits into direct sum of irreducible highest weight representations. There are two types of representations
\begin{equation}
    \text{NS}:\, r\in\mathbb{Z}+\frac{1}{2},\qquad
    \text{R}:\, r\in\mathbb{Z}.
\end{equation}
In the following we will consider NS representations.

Quantum $\mathcal{N}=2$ KdV system is a direct analog of the bosonic one \cite{Sasaki:1987mm}. 
It can be defined as an infinite dimensional abelian subalgebra in universal enveloping of $\mathcal{N}=2$ Virasoro algebra \eqref{N=2-Virasoro}. This subalgebra is spanned by local and quasilocal integrals of odd spins
\begin{equation}\label{IM-through-density}
    \mathbf{I}_{2k-1}=\frac{1}{2\pi}\int_{0}^{2\pi}G_{2k}(x)dx,\quad
    \tilde{\mathbf{I}}_{2k-1}=\frac{1}{2\pi}\int_{0}^{2\pi}\tilde{G}_{2k}(x)dx.
\end{equation}
Local densities are analytically regularized polynomials in currents \eqref{N=2-Virasoro-currents} and their derivatives. In particular
\begin{equation}
    G_2(x)=T(x)\implies \mathbf{I}_1=L_0-\frac{c}{24}.
\end{equation}
Conversely, the quasilocal densities (not to be confused to non-local ones) might include negative degrees of derivative (see below). 

In order to make contact with the supersymmetric $\mathfrak{sl}(2|1)$ Toda field theory, it is convenient to use the free field representation of the algebra \eqref{N=2-Virasoro}. Namely, the generating currents \eqref{N=2-Virasoro-currents} can be rewritten in terms of holomorphic free fields
\begin{equation}\label{XX-psipsi-OPE}
    X(z)X^*(w)=-2\log(z-w)+\dots,\quad\psi(z)\psi^*(w)=\frac{2}{z-w}+\dots
\end{equation}
as
\begin{equation}\label{N=2-generators}
\begin{gathered}
J=\frac{1}{2}\psi\psi^{*}-\frac{1}{b}\partial\bigl(X-X^{*}\bigr),\\
G^{+}=\frac{i}{2}\left(\partial X^{*}\psi-\frac{2}{b}\partial\psi\right),\quad
G^{-}=\frac{i}{2}\left(\partial X\psi^{*}-\frac{2}{b}\partial\psi^{*}\right),\\
T=-\frac{1}{2}\partial X\partial X^{*}-\frac{1}{4}\bigl(\psi\partial\psi^{*}+\psi^{*}\partial\psi\bigr)+\frac{1}{2b}\partial^{2}\bigl(X+X^{*}\bigr),
\end{gathered}
\end{equation}
where the Wick ordering is assumed. The modes of these currents obey the relations of $\mathcal{N}=2$ Virasoro algebra \eqref{N=2-Virasoro} with the central charge
\begin{equation}
    c=3\left(1+\dfrac{4}{b^{2}}\right).
\end{equation}
One can check that \eqref{N=2-generators} commute with the screening fields
\begin{equation}\label{screenings-CFT}
\mathcal{S}_1=\oint \psi^* e^{\frac{b}{2}X^*}dz,\qquad
\mathcal{S}_{2}=\oint \psi e^{\frac{b}{2}X}dz, 
\end{equation}
in the following sense
\begin{equation}
    \oint_{\mathcal{C}_z} \psi^*(\xi)e^{\frac{b}{2}X^*(\xi)}J(z)d\xi=
    \oint_{\mathcal{C}_z} \psi(\xi)e^{\frac{b}{2}X(\xi)}J(z)d\xi=0\quad\text{etc}.
\end{equation}

The screening operators $\mathcal{S}_1$ and $\mathcal{S}_2$ are the holomorphic counterparts of the exponential fields in the conformal action \eqref{SUSY Toda action} with $N=2$ and
\begin{equation}
    \boldsymbol{\alpha}_1=\frac{1}{2}(1,-i),\quad
    \boldsymbol{\alpha}_2=\frac{1}{2}(1,i)\implies
    \big(\boldsymbol{\alpha}_i\cdot\boldsymbol{\alpha}_j\big)=
    \begin{pmatrix}
        0&1\\
        1&0
    \end{pmatrix},
\end{equation}
which correspond to the superalgebra $\mathfrak{sl}(2|1)$. This theory is nothing else, but $\mathcal{N}=2$ Liouville theory. It is known to admit an integrable perturbation that correspond to $\mathcal{N}=2$ sinh-Gordon model, or affine $sl(2|2)$ Toda field theory in the terminology used above. On the level of screening fields one has to add
\begin{equation}\label{screenings-CFT-perturb}
\mathcal{S}_3=\oint \psi^* e^{-\frac{b}{2}X^*}dz,\qquad
\mathcal{S}_{4}=\oint \psi e^{-\frac{b}{2}X}dz. 
\end{equation}
The resulting theory corresponds to the affine root system $\widehat{\mathfrak{sl}}(2|2)$.

Quantum $\mathcal{N}=2$ KdV system is most naturally defined in terms of the free fields \eqref{N=2-generators}. Namely, the Integrals of Motion $\mathbf{I}_{2k-1}$ and $\tilde{\mathbf{I}}_{2k-1}$ are defined as quantities that commute with all the screening charges  $\mathcal{S}_k$ with $k=1,2,3,4$. The commutativity condition with $\mathcal{S}_1$ and $\mathcal{S}_2$ imply that the Integrals of Motion are local (or quasilocal) in terms of the basic currents \eqref{N=2-generators}, while commutativity with $\mathcal{S}_3$ and $\mathcal{S}_4$ is equivalent to the following condition for the corresponding densities \eqref{IM-through-density}
\begin{equation}\label{commutativity-equations}
\begin{aligned}
    \oint_{\mathcal{C}_z} \psi(\xi)e^{-\frac{b}{2}X(\xi)}G_{2k-1}(z)d\xi=\partial\mathcal{V}_{2k-1},\quad
    \oint_{\mathcal{C}_z} \psi^*(\xi)e^{-\frac{b}{2}X^*(\xi)}G_{2k-1}(z)d\xi=\partial\mathcal{V}^*_{2k-1},\\
    \oint_{\mathcal{C}_z} \psi(\xi)e^{-\frac{b}{2}X(\xi)}\tilde{G}_{2k-1}(z)d\xi=\partial\tilde{\mathcal{V}}_{2k-1},\quad
    \oint_{\mathcal{C}_z} \psi^*(\xi)e^{-\frac{b}{2}X^*(\xi)}\tilde{G}_{2k-1}(z)d\xi=\partial\tilde{\mathcal{V}}^*_{2k-1}.
\end{aligned}
\end{equation}
The set of equations \eqref{commutativity-equations} can be used to construct local and quasilocal Integrals of Motion explicitly. This calculation, being very tedious, nevertheless allows to obtain explicit expressions for IM's at lower levels.  The first two nontrivial representatives of both series have the form
\begin{equation}
    \tilde{\mathbf{I}}_1=2 \sum\limits_{n=1}^{\infty}J_{-n}J_n+2\sum\limits_{r>0}\frac{1}{r} \left(G_{-r}^{+}G_{r}^{-}+G_{-r}^-G_{r}^+\right),
\end{equation}
and
\begin{multline}
    \mathbf{I}_3=\sum_{r>0}\left(3(-1)^{r+\frac{1}{2}}+r\Big(9-\dfrac{12}{b^2}\Big)\right)\left(G_{-r}^+G^-_{r}+G^-_{-r}G^+_r\right)+
    \sum_{k=1}^{\infty}\left(\dfrac{3}{2}(-1)^{k+1}-\dfrac{c}{4}+\dfrac{k^2}{2}\Big(9-\dfrac{12}{b^2}\Big)\right)J_{-k}J_k-\\-\Big(\dfrac{3}{4}+\dfrac{c}{8} \Big)J_0^2+2\Big(9-\dfrac{12}{b^2}\Big)\sum_{k=1}^{\infty}L_{-k}L_k+\bigg(9-\dfrac{12}{b^2}\bigg)L_0^2+3\sum_{k=1}^{\infty}(L_{-(2k-1)}J_{2k-1}+J_{-(2k-1)}L_{2k-1})+\\+6\sum_{k+r+s=0}:J_{k}G^+_{r}G^-_{s}:+6\sum_{k+n+m=0}:L_{k}J_{n}J_{m}:+3L_0J_0^2+\dfrac{576(5+2c)-48b^2(39+20c)}{1152b^2}L_{0}+\\+\dfrac{-12c(13+2c)+b^2(123c+22c^2)}{1152b^2}.
\end{multline}
Here we write the Integrals of Motion in terms of modes. The corresponding expression for the densities can be found in appendix \ref{local-densities-app}. In principle, using the condition of commutativity with screening charges, one can compute higher Integrals of Motion.
\section{Spectrum and Bethe ansatz equations}\label{BAE-section}
In this section we consider the spectral problem for the quantum $\mathcal{N}=2$ $\text{KdV}_2$ system and provide the set of algebraic equations that describe the eigenvalues of Integrals of Motion (the Bethe ansatz equations). For simplicity we consider NS representation  of $\mathcal{N}=2$ Virasoro algebra spanned by the states
\begin{equation}
   \mathcal{V}^{\scriptscriptstyle{\textrm{NS}}}_{\Delta,q}=\text{span}
    \left(L_{-\boldsymbol{\lambda}}J_{-\boldsymbol{\mu}}G^+_{-\boldsymbol{r}}G^{-}_{-\boldsymbol{s}}|\Delta,q\rangle\right)
    \quad L_{-\boldsymbol{\lambda}}
    \overset{\text{def}}{=}L_{-\lambda_1}L_{-\lambda_2}\dots\quad\text{for}\quad
    \boldsymbol{\lambda}=(\lambda_1\geq\lambda_2\geq\dots)\quad\text{etc}
\end{equation}
where the highest weight state $|\Delta,q\rangle$ is defined as follows
\begin{equation}
\begin{gathered}
L_n|\Delta,q\rangle=J_n|\Delta,q\rangle=G^{\pm}_r|\Delta,q\rangle=0
\quad\text{for}\quad n,r>0,\\
L_0|\Delta,q\rangle=\Delta|\Delta,q\rangle,\quad
J_0|\Delta,q\rangle=q|\Delta,q\rangle.
\end{gathered}
\end{equation}
For generic values of $\Delta$ and $q$ the representation $\mathcal{V}^{\scriptscriptstyle{\textrm{NS}}}_{\Delta,q}$ is irreducible. It can be decomposed into direct sum of finite dimensional subspaces with the fixed eigenvalues of the operators $L_0$ and $J_0$
\begin{equation}
   \mathcal{V}^{\scriptscriptstyle{\textrm{NS}}}_{\Delta,q}=
   \bigoplus_{N_1=0}^{\infty}\bigoplus_{N_2=0}^{\infty}
   \mathcal{V}^{\scriptscriptstyle{\textrm{NS}},N_1,N_2}_{\Delta,q},
\end{equation}
where
\begin{equation}
\mathcal{V}^{\scriptscriptstyle{\textrm{NS}},N_1,N_2}_{\Delta,q}=
\text{span}\left[|k\rangle\in\mathcal{V}^{\scriptscriptstyle{\textrm{NS}}}_{\Delta,q}:
L_0|k\rangle=\left(\Delta+\frac{1}{2}\big(N_1+N_2\big)\right)|k\rangle,\,
J_0|k\rangle=\big(N_1-N_2\big)|k\rangle\right]
\end{equation}
The number of states in $\mathcal{V}^{\scriptscriptstyle{\textrm{NS}},N_1,N_2}_{\Delta,q}$ can be computed from the generating function
\begin{equation}\label{character}
    \chi(q,t)\overset{\text{def}}{=}\sum_{N_1=0}^{\infty}\sum_{N_2=0}^{\infty}
    p(N_1,N_2)q^{\frac{N_1+N_2}{2}}t^{N_1-N_2}
    =\prod_{k=1}^{\infty}\frac{\left(1+tq^{k-\frac{1}{2}}\right)\left(1+t^{-1}q^{k-\frac{1}{2}}\right)}{(1-q^k)^2}.
\end{equation}

The Integrals of Motion $\mathbf{I}_{2k-1}$ and $\tilde{\mathbf{I}}_{2k-1}$ commute with the zero modes $L_0$ and $J_0$. It implies that they can be diagonalized in $\mathcal{V}^{\scriptscriptstyle{\textrm{NS}},N_1,N_2}_{\Delta,q}$. For given $N_1$ and $N_2$ $\mathbf{I}_{2k-1}$ and $\tilde{\mathbf{I}}_{2k-1}$ reduce to finite matrices. Let us fix the pair $N_1$ and  $N_2$ and associate to it the set of Bethe roots 
\begin{equation}
    x_1,\dots,x_{N_1}\quad\text{and}\quad y_1,\dots,y_{N_2},
\end{equation}
and the functions
\begin{equation}\label{BAE-functions}
    S(x)=\frac{x^2-h_1^2}{x^2-h_2^2},\quad
    A(x)=\frac{\left(x-\frac{h_2}{2}\right)^2-\left(\frac{u-u^*}{2}\right)^2}
    {\left(x-\frac{h_1}{2}\right)^2-\left(\frac{u+u^*}{2}\right)^2},\quad
    B(x)=\frac{\left(x+\frac{h_1}{2}\right)^2-\left(\frac{u+u^*}{2}\right)^2}
    {\left(x+\frac{h_2}{2}\right)^2-\left(\frac{u-u^*}{2}\right)^2},
\end{equation}
where
\begin{equation}\label{hh-b}
    h_1=\frac{b}{2}+b^{-1},\quad h_2=\frac{b}{2}-b^{-1}.
\end{equation}
Our main result states that the following Bethe ansatz equations
\begin{equation}\label{BAE}
    \begin{aligned}
        &A(x_i)\prod_{j=1}^{N_2}S(x_i-y_j)=1,\quad&&\text{for}\quad i=1,\dots,N_1,\\
        &B(y_i)\prod_{j=1}^{N_1}S^{-1}(y_i-x_j)=1,\quad&&\text{for}\quad i=1,\dots,N_2,
    \end{aligned}
\end{equation}
describe the spectrum of quantum $\mathcal{N}=2$ $\text{KdV}_2$ system with the central charge $c$ and the highest weight parameters $\Delta$ and $q$ given by
\begin{equation}
    c=3 \left(1 + \frac{4}{b^2}\right)\quad\text{and}\quad
    q=\frac{u-u^*}{b},\quad\Delta=\frac{1}{2b^2}-\frac{uu^*}{2}.
\end{equation}
The eigenvalues of local and semi-local integrals of motion are symmetric polynomials in Bethe roots. In particular, we have found that
\begin{equation}\label{Itilde-x}
    \tilde{\mathbf{I}}_1\sim 8b^{-1}\left(\sum_{i=1}^{N_1}x_i-b\big(N_1-N_2\big)\left(\frac{\Delta+N_1}{2}+\frac{q}{4}\right)+\frac{N_1(2+b^2)}{4b}\right),
\end{equation}
where the sign $\sim$ means that the corresponding eigenvalues are given by the expression in the right hand side with Bethe roots solving \eqref{BAE}. 

One might wounder that the expression \eqref{Itilde-x} does include the roots $x_k$, but not $y_k$. However, it is well known that symmetric polynomials of Bethe roots are algebraically related. In particular, based on explicit calculations on lower levels we have empirically established the constraint
\begin{equation}
    \sum_{i=1}^{N_1}x_i+\sum_{j=1}^{N_2}y_j=-\frac{b}{4}\big(N_1-N_2\big)
    \Big(2uu^*-2\big(N_1+N_2\big)+1\Big).
\end{equation}

The fact that the spectrum of $\mathcal{N}=2$ $\text{KdV}_2$ system is described by Bethe equations of the form \eqref{BAE} implicitly follow from the results of \cite{Kolyaskin:2022tqi}. Namely, in \cite{Kolyaskin:2022tqi} it has been show that $\mathcal{N}=2$ $W_n$ algebras appear naturally as representations of affine Yangian of $\widehat{\mathfrak{gl}}(1|1)$ (it has been already noticed in \cite{Prochazka:2015deb,Gaberdiel:2017dbk,Gaberdiel:2017hcn,Gaberdiel:2018nbs}). This fact provides natural definition of the transfer-matrix whose expansion at large values of the spectral parameter gives local Integrals of Motion of some integrable system. It can be shown, that in the case of untwisted transfer-matrix this integrable system coincides with $\mathcal{N}=2$ $\text{KdV}_n$ system for a class of representations of $\textrm{Y}\left(\widehat{\mathfrak{gl}}(1|1)\right)$ on $n$ sites. The spectrum of this integrable system is governed by Bethe ansatz equations that can be derived using a version of algebraic Bethe ansatz approach developed in \cite{Litvinov:2020zeq,Litvinov:2021phc,Chistyakova:2021yyd}. In the next section we will outline this derivation.
\section{Affine Yangian $Y(\widehat{\mathfrak{gl}}(1|1))$ and Bethe ansatz equations}\label{Ygl(1|1)-ansatz}
The commutation relations of $Y(\widehat{\mathfrak{gl}}(1|1))$ can be compactly written in terms of modes of the currents admitting the expansion at $\infty$
\begin{equation}
    e_k(u)=\frac{e_k^{(0)}}{u}+\frac{e_k^{(1)}}{u^2}+\dots,\quad
    f_k(u)=\frac{e_k^{(0)}}{u}+\frac{e_k^{(1)}}{u^2}+\dots,\quad
    \uppsi_k(u)=\frac{e_k^{(0)}}{u}+\frac{e_k^{(1)}}{u^2}+\dots,\quad
    k=1,2,
\end{equation}
as
\begin{equation}\label{Yangian-relation-1}
\left[\uppsi_i(u),\uppsi_j(v)\right]=0,\quad
\left\{e_i(u),e_i(v)\right\}=0,\quad\left\{f_i(u),f_i(v)\right\}=0,
\end{equation}
\begin{equation}\label{ef+fe}
\left\{e_i(u),f_j(v)\right\}=\delta_{ij}\frac{\uppsi_i(u)-\uppsi_i(v)}{u-v},
\end{equation}
\begin{equation}\label{ee-relation}
e_1(u)e_2(v)=-\frac{(v-u-h_2)(v-u+h_2)}{(u-v-h_1)(u-v+h_1)}e_2(v)e_1(u)+\text{\textcolor{blue}{local}},
\end{equation}
\begin{equation}\label{ff-relation}
f_2(u)f_1(v)=-\frac{(v-u-h_2)(v-u+h_2)}{(u-v-h_1)(u-v+h_1)}f_1(v)f_2(u)+\text{\textcolor{blue}{local}},
\end{equation}
\begin{equation}\label{psie-relation}
\uppsi_i(u)e_j(v)=\frac{(u-v-h_j)(u-v+h_j)}{(u-v-h_i)(u-v+h_i)}e_j(v)\uppsi_i(u)+\text{\textcolor{blue}{local}},
\end{equation}
\begin{equation}\label{psif-relation}
\uppsi_i(u)f_j(v)=\frac{(u-v-h_i)(u-v+h_i)}{(u-v-h_j)(u-v+h_j)}f_j(v)\uppsi_i(u)+\text{\textcolor{blue}{local}},
\end{equation}
where by \textcolor{blue}{local} one denotes operators that depend only on one variable either $u$ or $v$ (see \cite{Kolyaskin:2022tqi}). These terms do not contribute to the commutation relations of modes with indexes higher than $1$. Note that despite the similarity of the relations \eqref{ee-relation}--\eqref{psif-relation} to OPE in CFT, the product of the currents is regular. So the visible poles in the relations \eqref{ee-relation}--\eqref{psif-relation} are canceled by the local terms.

Affine Yangian of $\mathfrak{gl}(1|1)$ appears naturally in conformal field theory as a subalgebra  of $RLL$ algebra. Namely consider the following Lax operator \cite{Kolyaskin:2022tqi}
\begin{equation}\label{Lax-definition}
    \mathcal{L}^{(0)}=-b^{-2}\partial+\frac{1}{2}\left(b^{-1}\partial X-\psi\psi^*-b^{-1}\partial X^*\right)+
    \frac{1}{2b}\left(\partial\psi-b\psi\partial X^*\right)\theta
    -b^{-1}\psi^*\frac{\partial}{\partial\theta}+b^{-1}\partial X^*\theta\frac{\partial}{\partial\theta},
\end{equation}
where $(X,X^*,\psi,\psi^*)$ are the fields satisfying \eqref{XX-psipsi-OPE}, but in different normalization, with the factor $1$ instead of $2$ in the right hand side. For simplicity we consider NS representation for fermions, i.e. in terms of the modes on the cylinder $z=e^{-ix}$, $x\sim x+2\pi$ one has
\begin{equation}
\begin{gathered}
    \partial X(x)=\sum_{k\in\mathbb{Z}}X_ke^{-ikx},\quad
    \partial X^*(x)=\sum_{k\in\mathbb{Z}}X^*_ke^{-ikx},\quad
    \psi(x)=i^{\frac{1}{2}}\sum_{r\in\mathbb{Z}+\frac{1}{2}}\psi_re^{-irx},\quad
    \psi^*(x)=i^{\frac{1}{2}}\sum_{r\in\mathbb{Z}+\frac{1}{2}}\psi^*_re^{-irx},
\end{gathered}
\end{equation}
with commutation relations (other commutation relations are trivial)
\begin{equation}
    [X_m,X_n^*]=m\delta_{m,-n},\quad\{\psi_{r},\psi_{s}^*\}=\delta_{r,-s}.
\end{equation}
Then the Fock module $\mathcal{F}_{\boldsymbol{u}}$, where $\boldsymbol{u}=(u,u^*)$, is generated from  the vacuum state $|\boldsymbol{u}\rangle$
\begin{equation}
\begin{gathered}
    a_{n}|\boldsymbol{u}\rangle=a^*_{n}|\boldsymbol{u}\rangle=\psi_{r}|\boldsymbol{u}\rangle=
    \psi^*_{r}|\boldsymbol{u}\rangle=0\quad\text{for}\quad n>0,\,r>0,\\
    a_{0}|\boldsymbol{u}\rangle=-iu|\boldsymbol{u}\rangle,\quad
    a_{0}^*|\boldsymbol{u}\rangle=-iu^*|\boldsymbol{u}\rangle,
\end{gathered}
\end{equation}
by the action of the creation operators
\begin{equation}\label{NS-module}
    \mathcal{F}_{\boldsymbol{u}}=\textrm{Span}\left(\psi_{-\boldsymbol{s}^*}^*\psi_{-\boldsymbol{s}}a^*_{-\boldsymbol{\lambda}^*}a_{-\boldsymbol{\lambda}}|\boldsymbol{u}\rangle\right),\quad\text{where}\quad
    a_{-\boldsymbol{\lambda}}=a_{-\lambda_1}a_{-\lambda_2}\dots\,\text{etc}.,
\end{equation}
where
\begin{equation}
    \boldsymbol{\lambda}=\{\lambda_1\geq\lambda_2\geq\dots\},\quad
    \boldsymbol{\lambda}^*=\{\lambda_1^*\geq\lambda_2^*\geq\dots\},\quad
    \boldsymbol{s}=\{s_1>s_2>\dots\},\quad
    \boldsymbol{s}^*=\{s_1^*>s_2^*>\dots\}.
\end{equation}
Thus in given representation one has $\mathcal{L}=\mathcal{L}(\boldsymbol{u})$. The Fock representation $\mathcal{F}_{\boldsymbol{u}}$ admits the action  of the current algebra $\widehat{\mathfrak{gl}}(1|1)$ in the so-called typical representation. The components of the Lax operator \eqref{Lax-definition} correspond to the current $\boldsymbol{E}\in\widehat{\mathfrak{gl}}(1|1)$ as
\begin{equation}
    \boldsymbol{E}=
    \begin{pmatrix}
      b^{-1}\partial X-\psi\psi^*&\psi^*\\
      \partial\psi-b\psi\partial X^*&\psi\psi^*-b^{-1}\partial X-b\partial X^*
    \end{pmatrix}.
\end{equation}

Having defined the Lax operator \eqref{Lax-definition}, the $R-$matrix of $\textrm{Y}(\widehat{\mathfrak{gl}}(1|1))$ is defined as an intertwining operator acting in the tensor product of two Fock modules $\mathcal{F}_{\boldsymbol{u}_1}$  and $\mathcal{F}_{\boldsymbol{u}_2}$
\begin{equation}\label{RLL-Miura}
\mathcal{R}_{12}(\boldsymbol{u}_1|\boldsymbol{u}_2)\mathcal{L}_1^{(0)}(\boldsymbol{u}_1)\mathcal{L}^{(0)}_2(\boldsymbol{u}_2)=
\mathcal{L}_2^{(0)}(\boldsymbol{u}_2)\mathcal{L}_1^{(0)}(\boldsymbol{u}_1)\mathcal{R}_{12}(\boldsymbol{u}_1|\boldsymbol{u}_2),
\end{equation}
where by $\mathcal{L}_k^{(0)}(\boldsymbol{u}_k)$ one denotes \eqref{Lax-definition} with $X(z)\rightarrow X_k(z)$. Using \eqref{RLL-Miura} one can compute the matrix of the operator $\mathcal{R}_{12}(\boldsymbol{u}_1|\boldsymbol{u}_2)$ at any given level, but the computations become very complicated as the level grows (see \cite{Kolyaskin:2022tqi} for details). We note that \eqref{RLL-Miura} is the special case of $RLL$ algebra with the Lax operator $\mathcal{L}^{(0)}(\boldsymbol{u})$ whose quantum and auxiliary spaces (in the standard terminology of the spin chains) are
\begin{equation}
    \begin{aligned}
        &\text{quantum space}:\quad&&\text{functions of $x$ and $\theta$, where $\partial$ and $\frac{\partial}{\partial\theta}$ act},\\
        &\text{auxiliary space}:\quad&&\text{Fock module}\quad\mathcal{F}_{\boldsymbol{u}}.
    \end{aligned}
\end{equation}
On the other hand with quantum space being unspecified  one has the algebra
\begin{equation}\label{RLL-general}
\mathcal{R}_{12}(\boldsymbol{u}_1|\boldsymbol{u}_2)\mathcal{L}_1(\boldsymbol{u}_1)\mathcal{L}_2(\boldsymbol{u}_2)=
\mathcal{L}_2(\boldsymbol{u}_2)\mathcal{L}_1(\boldsymbol{u}_1)\mathcal{R}_{12}(\boldsymbol{u}_1|\boldsymbol{u}_2),
\end{equation}
which we call $RLL$ or the Yang-Baxter algebra of $\widehat{\mathfrak{gl}}(1|1)$. It has been shown in \cite{Kolyaskin:2022tqi} that \eqref{RLL-general} contains $Y(\widehat{\mathfrak{gl}}(1|1))$ as subalgebra with parameters $h_k$ given by \eqref{hh-b}. 

Apart from  $e_k(u)$, $f_k(u)$ and $\uppsi_k(u)$ the algebra \eqref{RLL-general} also contains auxiliary Cartan current $h(\boldsymbol{u})=h(u,u^*)$ that depend on two spectral parameters rather than one.  This current satisfies the following commutation relations with $\textrm{Y}(\widehat{\mathfrak{gl}}(1|1))$ currents \cite{Kolyaskin:2022tqi}
\begin{equation}
[h(\boldsymbol{u}),h(\boldsymbol{v})]=[h(\boldsymbol{u}),\uppsi_k(v)]=0.
\end{equation}
\begin{equation}\label{he-relation}
h(\boldsymbol{u})e_1(v)=\frac{(u-u^*)-v}{(u+u^*)-v+b^{-1}}\,e_1(v)h(\boldsymbol{u})+\text{\textcolor{blue}{local}},
\end{equation}
\begin{equation}\label{he*-relation}
h(\boldsymbol{u})e_2(v)=\frac{(u+u^*)-v-b/2}{(u-u^*)-v-h_2}\,e_2(v)h(\boldsymbol{u})+\text{\textcolor{blue}{local}},
\end{equation}
\begin{equation}\label{hf*-relation}
f_1(v)h(\boldsymbol{u})=\frac{(u-u^*)-v}{(u+u^*)-v+b^{-1}}\,h(\boldsymbol{u})f_1(v)+\text{\textcolor{blue}{local}}.
\end{equation}
\begin{equation}\label{hf-relation}
f_2(v)h(\boldsymbol{u})=
\frac{(u+u^*)-v-b/2}{(u-u^*)-v-h_2}\,h(\boldsymbol{u})f_2(v)+\text{\textcolor{blue}{local}}.
\end{equation}

Different representations of \eqref{RLL-general} correspond to different CFT's. We will consider the most natural representation
\begin{equation}
    \mathcal{L}(\boldsymbol{u})=\mathcal{R}_{01}(\mathbf{u}|\mathbf{u}_1)\dots
    \mathcal{R}_{0n}(\mathbf{u}|\mathbf{u}_n),
\end{equation}
which can be thought as a generalization of the  spin-chain representation on $n-$sites. The corresponding transfer matrix has the form
\begin{equation}\label{T-prime}
    \boldsymbol{T}(\boldsymbol{u})=\textrm{Tr}'\left(\mathcal{R}_{01}(\mathbf{u}|\mathbf{u}_1)\dots
    \mathcal{R}_{0n}(\mathbf{u}|\mathbf{u}_n)\right),
\end{equation}
where $'$ means certain regularization of the trace (see \cite{Litvinov:2020zeq,Chistyakova:2021yyd} for details). This transfer matrix corresponds to quantum $\mathcal{N}=2$ $\textrm{KdV}_n$ system in the certain limit. In order to see it, let us consider the case $n=2$. In this case an immediate consequence of \eqref{RLL-general} shows
\begin{equation}
      [\boldsymbol{T}(u),\mathcal{R}_{12}(\boldsymbol{u}_1|\boldsymbol{u}_2)]=0\quad\text{for}\quad
      n=2.
\end{equation}
The operator $\mathcal{R}_{12}(\boldsymbol{u}_1|\boldsymbol{u}_2)$ serves as an intertwining operator \eqref{RLL-Miura} between two $W-$algebras.  It has been shown in \cite{Kolyaskin:2022tqi} that the currents appearing in the product $\mathcal{L}_1^{(0)}\cdot\mathcal{L}_2^{(0)}$  can be defined as a commutant of the set of screening fields
\begin{equation}\label{Isomorphic screenings_1}
\hat{\mathcal{S}}_1=\int\psi_1^*e^{bX_1^*}dz\quad\hat{\mathcal{S}}_{1,2}=\int (\psi_1-\psi_2)e^{\frac{b}{2}\left(X_1-X_2-X_1^*-X_2^*\right)}dz, \quad \hat{\mathcal{S}}_2=\int\psi_2^*e^{bX_2^*}dz
\end{equation}
There are eight currents, including $\left(J_{12},G_{12},G_{12}^*,T_{12}\right)$, that form $\mathcal{N}=2$ Virasoro algebra with the central charge $c=6.$ The screenings $\mathcal{S}_1$, $\mathcal{S}_2$ are the Wakimoto type operators that define two copies of $\widehat{\mathfrak{gl}}(1|1)$ current algebra, and $\mathcal{S}_{1,2}$ provides the interaction between the two.  The integrable perturbation of this model is implemented by the additional screening field \cite{Kolyaskin:2022tqi}
\begin{equation}\label{Affine screening}
 \hat{\mathcal{S}}_{2,1}=\int (\psi_1-\psi_2)e^{\frac{b}{2}\left(X_2-X_1-X_1^*-X_2^*\right)}dz.
\end{equation}
If we drop this exponent, the symmetry algebra of the initial theory with the screenings ($\mathcal{S}_1$, $\mathcal{S}_{1,2}$, $\mathcal{S}_2$) is described by the Lax operator $\mathcal{L}_1^{(0)}\cdot\mathcal{L}_2^{(0)}$. On the other hand, we can consider different but isomorphic $W-$algebra defined by the set of screenings ($\mathcal{S}_2$, $\mathcal{S}_{2,1}$, $\mathcal{S}_1$) with the Lax operator $\mathcal{L}_2^{(0)}\cdot\mathcal{L}_1^{(0)}$, which are connected by $RLL$-algebra (\ref{RLL-Miura}). It is clear, that the Integrals of Motion should belong to the intersection of these two $W$-algebras and commute with $\mathcal{R}_{12}(\boldsymbol{u}_1|\boldsymbol{u}_2)$. 

In order to see how the integrable system provided by the screening fields \eqref{Isomorphic screenings_1} and \eqref{Affine screening} is related to $\mathcal{N}=2$ quantum $\text{KdV}_2$ hierarchy it is convenient to transform to new coordinates. Namely, we introduce complex bosons ($X,X^*,\widetilde{X},\widetilde{X}^*$) as
\begin{equation} \label{bosons}
X_1=\dfrac{X+\lambda^{-1}\widetilde{X}}{2}, \quad X_2=\dfrac{-X+\lambda^{-1}\widetilde{X}}{2}, \quad     X_1^*=\dfrac{X^*+\lambda\widetilde{X}^*}{2}, \quad X_2^*=\dfrac{-X^*+\lambda\widetilde{X}^*}{2},
\end{equation}
and complex fermions ($\psi,\psi^*,\widetilde{\psi},\widetilde{\psi}^*$):
\begin{equation}\label{fermions}
\psi_1=\dfrac{\psi+\lambda^{-1}\widetilde{\psi}}{2}, \quad \psi_2=\dfrac{-\psi+\lambda^{-1}\widetilde{\psi}}{2}, \quad   \psi_1^*=\dfrac{\psi^*+\lambda\widetilde{\psi}^*}{2}, \quad \psi_2^*=\dfrac{-\psi^*+\lambda\widetilde{\psi}^*}{2}.
\end{equation}
Here  $\lambda$ is a formal parameter that does not alter commutation relations of the fields $(\tilde{X},\tilde{X}^*,\tilde{\psi},\tilde{\psi}^*)$. We notice that in the limit $\lambda\rightarrow 0$ these fields  decouple from \eqref{Isomorphic screenings_1} and \eqref{Affine screening} and the resulting screening fields take the form
\begin{equation}\label{fermionic-screenings}
\mathcal{S}_1=\oint \psi^* e^{\frac{b}{2}X^*}dz,\qquad
\mathcal{S}_{2}=\oint \psi e^{\frac{b}{2}X}dz, \qquad
\mathcal{S}_3=\oint \psi^* e^{-\frac{b}{2}X^*}dz,\qquad
\mathcal{S}_{4}=\oint \psi e^{-\frac{b}{2}X}dz, 
\end{equation}
that is precisely the one for $\mathcal{N}=2$ $\text{KdV}_2$ systems \eqref{screenings-CFT} and \eqref{screenings-CFT-perturb}. Similar logic allows to decouple the "center of mass" degrees of freedom for the theory with $n-$degrees of freedom.

Having defined the integrable system, Bethe ansatz equations for its spectrum follow from the results of \cite{Litvinov:2020zeq,Chistyakova:2021yyd}.  We comment on the simplest case $n=2$. The key idea is to diagonalize the operator $\mathcal{R}_{12}(\boldsymbol{u}_1|\boldsymbol{u}_2)$ instead of local Integrals of Motion. In order to do this one defines the off-shell Bethe vector
\begin{equation}\label{Bethe-vector}
    |B(\boldsymbol{x},\boldsymbol{y})\rangle\in\mathcal{F}_{\boldsymbol{u}_1}\otimes
    \mathcal{F}_{\boldsymbol{u}_2}.
\end{equation}
The  vector $|B(\boldsymbol{x},\boldsymbol{y})\rangle$ has been constructed in such a way that its projection on arbitrary state $|\psi_1\rangle\otimes|\psi_2\rangle=|\psi_1,\psi_2\rangle$ can be expressed as a matrix element in the auxiliary space
\begin{equation}\label{B-projection-1}
    \langle\psi_1,\psi_2|B(\boldsymbol{x},\boldsymbol{y})\rangle=
    \,_
    {\scriptscriptstyle{\text{aux}}}\langle\varnothing|
    \mathcal{L}_{\psi_1,\varnothing}(\boldsymbol{u}_1)
    \mathcal{L}_{\psi_2,\varnothing}(\boldsymbol{u}_2)
    |\chi(\boldsymbol{x},\boldsymbol{y})\rangle_
    {\scriptscriptstyle{\text{aux}}},
\end{equation}
where $\mathcal{L}_{\psi,\varnothing}(\boldsymbol{u})=\langle\psi|\mathcal{L}(\boldsymbol{u})|\boldsymbol{u}\rangle$ is the  element of the matrix $\mathcal{L}(\boldsymbol{u})$ and $|\chi(\boldsymbol{x},\boldsymbol{y})\rangle_
    {\scriptscriptstyle{\text{aux}}}$ is the special state in the auxiliary space which depends on variables $\boldsymbol{x}=(x_1,\dots,x_{N_1})$ and $\boldsymbol{y}=(y_1,\dots,y_{N_2})$. In these terms the action of the operator $\mathcal{R}_{12}(\boldsymbol{u}_1|\boldsymbol{u}_2)$ is very simple
\begin{equation}\label{B-projection-2}
    \langle\psi_1,\psi_2|\mathcal{R}_{12}(\boldsymbol{u}_1|\boldsymbol{u}_2)|B(\boldsymbol{x},\boldsymbol{y})\rangle=
    \,_
    {\scriptscriptstyle{\text{aux}}}\langle\varnothing|
    \mathcal{L}_{\psi_2,\varnothing}(\boldsymbol{u}_2)
    \mathcal{L}_{\psi_1,\varnothing}(\boldsymbol{u}_1)
    |\chi(\boldsymbol{x},\boldsymbol{y})\rangle_
    {\scriptscriptstyle{\text{aux}}},
\end{equation}
In order to compare \eqref{B-projection-1} and \eqref{B-projection-2} it is useful to know how to commute $\mathcal{L}_{\psi_2,\varnothing}(\boldsymbol{u}_2)$ and $\mathcal{L}_{\psi_1,\varnothing}(\boldsymbol{u}_1)$.  It is  done using the fact that $\mathcal{L}_{\psi,\varnothing}(\boldsymbol{u})$ can be represented by the contour integral \cite{Litvinov:2020zeq,Chistyakova:2021yyd}
\begin{equation}\label{L-integral}
    \mathcal{L}_{\psi,\varnothing}(\boldsymbol{u})=\oint F_{\psi}(\boldsymbol{z},\boldsymbol{w})f_1(z_1)\dots f_1(z_{n_1})f_2(w_1)\dots f_2(w_{n_2})h(\boldsymbol{u})d\boldsymbol{z}d\boldsymbol{w},
\end{equation}
where $n_1$ and $n_2$ are related to conformal dimension and charge of the states $|\psi\rangle$. Integration in \eqref{L-integral} goes around infinity and includes all
singularities of the rational function $F_{\psi}(\boldsymbol{z},\boldsymbol{w})$ which encodes  information about the state $|\psi\rangle$. The explicit form of this function is important for actual form of the Bethe vector \eqref{Bethe-vector}, but not for Bethe ansatz equations. Using the Cauchy theorem the integral in \eqref{L-integral} localizes to the poles coming from the state $|\chi(\boldsymbol{x},\boldsymbol{y})\rangle_{\scriptscriptstyle{\text{aux}}}$, the position of these poles are the Bethe roots. On the other hand permutation of two $\mathcal{L}(\boldsymbol{u})$ operators in \eqref{B-projection-2}  will produce extra factors coming from the commutation relations \eqref{hf*-relation}, \eqref{hf-relation} and \eqref{ff-relation}. These factors are nothing but the functions $S(x)$, $A(x)$ and $B(x)$ in \eqref{BAE-functions} after the scaling provided by \eqref{bosons} and in the limit $\lambda\rightarrow0$. 

We should stress that while the derivation of Bethe ansatz equations is rather straightforward and follows exactly \cite{Litvinov:2020zeq,Chistyakova:2021yyd}, the problem of expressing the eigenvalues of local and semi-local IM's through the Bethe roots is tough (see \cite{Litvinov:2020zeq}). Of course the eigenvalues are some symmetric functions of the roots. Moreover, from explicit examples \cite{Litvinov:2013zda,Alfimov:2014qua} it is known that they are in fact symmetric polynomials.  Admitting this conjecture in current case we have found \eqref{Itilde-x} by direct diagonalization and comparing to the solution of BAE up to level $4$. Moreover \eqref{Itilde-x} can be considered as an independent of of the whole machinery used in \cite{Litvinov:2013zda,Alfimov:2014qua,Kolyaskin:2022tqi}.
\section{Concluding remarks}
In these notes we have found Bethe ansatz equations \eqref{BAE} for the spectrum of $\mathcal{N}=2$ KdV system. Equations \eqref{BAE} constitute main our main results. Here we provide some remarks and open questions.
\begin{itemize}
       \item Our results admits straightforward  generalization for $\mathcal{N}=2$ $\text{KdV}_n$ system with $n>2$ which correspond to integrable perturbations of Kazama-Suzuki models \cite{Kazama:1988qp}. The Bethe ansatz equations in this case are given by \eqref{BAE}, but with with the source terms $A(x)$ and $B(x)$ being the ratio of polynomials of degree $n$ (see \cite{Litvinov:2020zeq,Chistyakova:2021yyd}).
    \item The approach to integrability in CFT based on affine Yangian symmetry naturally defines twisted integrable system \cite{Litvinov:2021phc,Chistyakova:2021yyd}. Namely, instead of regularized transfer-matrix \eqref{T-prime}  one considers twisted transfer matrix
    \begin{equation}\label{T-twisted}
    \boldsymbol{T}(\boldsymbol{u})=\textrm{Tr}\left(q^{L_0^{(0)}}t^{h_0^{(0)}}\mathcal{R}_{01}(\mathbf{u}|\mathbf{u}_1)\dots
    \mathcal{R}_{0n}(\mathbf{u}|\mathbf{u}_n)\right),
\end{equation}
where $L_0^{(0)}$ and $h_0^{(0)}$ act in the auxiliary space (here $h_0$ is the zero mode of $\psi\psi^*$). It would be interesting to identify Toda field theory that corresponds to this integrable system.
    \item The quantum $\mathcal{N}=2$ $\text{KdV}_2$ system is a particular case of the Fateev model \cite{Fateev:1996ea} (see also \cite{Lukyanov:2013wra}). This model is purely bosonic and corresponds to the exceptional root system $\mathrm{D}(2,1|\alpha)$. In particular, its screening operators do not include fermions. They can be obtained from \eqref{fermionic-screenings} by boson-fermion correspondence. Thus our BA equations provide the spectrum of the Fateev model in the supersymmetric point. However, we do not expect that the full Fateev model is related to $\mathrm{Y}(\widehat{\mathfrak{gl}}(1|1))$. It rather corresponds to $\mathrm{Y}(\widehat{\mathfrak{gl}}(1))$ with the "boundary" \cite{Litvinov:2021phc}, but it requires further study.  In particular, it is still a challenging problem to find BAE in this case. 
\end{itemize}
\section*{Acknowledgments}
This work has been supported by the Russian Science Foundation under the grant 22-22-00991.
\appendix
\section{Quantum $\mathcal{N}=2$ KdV integrals and analytic ordering}\label{local-densities-app}
In this appendix we will comment on construction of quantum $\mathcal{N}=2$ KdV integrals and on analytic regularization of the densities. We will work in $\mathcal{N}=2$ notations. We define the holomorphic part of supercoordinates $Z=(z,\theta, \bar{\theta}).$ and introduce the holomorphic superfield
\begin{equation}
\mathbb{T}=J+\theta^+ G^+ + \theta^- G^-+\theta^+ \theta^-T
\end{equation}
and super derivatives
\begin{equation}
D^+=\frac{\partial}{\partial\theta^+}-\dfrac{1}{2}\theta^-\partial, \qquad D^-=\frac{\partial}{\partial\theta^-}-\dfrac{1}{2}\theta^+\partial.
\end{equation}
Then in these notations the Integrals of Motion can be written in supersymmetric form and satisfy quantum $\mathcal{N}=2$ KdV hierarchy
\begin{equation}\label{IMs}
\begin{aligned}
&\mathbf{I}_{1}=\int dz d\theta^+ d\theta^-\mathbb{T},
\\
&\mathbf{I}_{3}=\int dz d\theta^+ d\theta^-\left((\mathbb{T}(\mathbb{T}\mathbb{T}))+\left( 9-\frac{12}{b^2}\right)(D^+\mathbb{T}D^-\mathbb{T})\right),
\\
&\dots
\end{aligned}
\end{equation}
where by $(AB)$ we denoted analytic ordering of two operators $A,$ $B$ which is defined as follows. Consider two local operators $A(z)$ and $B(z)$ 
\begin{equation}
    A(z)=\sum_{r}A_r e^{-i r z}, \quad B(z)=\sum_{r}B_r e^{-i r z},
\end{equation}
where we assume that the fields can be either bosonic ($r\in\mathbb{Z}$) or fermionic in NS sector ($r\in\mathbb{Z}+\frac{1}{2}$). Analytic ordering is defined as follows
\begin{equation}
    (A(z) B(z)) \stackrel{\text { def }}{=} \frac{1}{2 \pi i} \oint_{\mathcal{C}_z} \frac{\mathcal{T}(A(w) B(z))}{w-z} d w=\sum_{r}\Lambda_r e^{-i r z},
\end{equation}
where summation goes over $r\in \mathbb{Z}$ if both currents are bosonic or fermionic, and $r\in \mathbb{Z}+\frac{1}{2}$ otherwise. $\mathcal{T}$ stands for chronological order 
\begin{equation}
\mathcal{T}(A(w) B(z))= \begin{cases}A(w) B(z), & \quad \Im(w)<\Im(z), 
\\ B(z) A(w), & \quad   \Im(w)>\Im(z),\end{cases}
\end{equation}
and by definition
\begin{equation}\label{Lambda}
    \Lambda_r\stackrel{\text { def }}{=}\frac{1}{2 \pi} \int_0^{2 \pi} e^{i r z} ( A(z) B(z)) d z=\frac{1}{4 i \pi^2} \int_0^{2 \pi} \oint_{\mathcal{C}_z} e^{i r z} \frac{\mathcal{T}(A(w) B(z))}{w-z} d w d z.
\end{equation}
For what follows, it will be convenient to "approximate" $\frac{1}{z}$ by periodic 
\begin{equation}\label{Approximation_1}
    \frac{1}{z}=\frac{1}{2}\cot\frac{z}{2}-2\sum_{q=1}^{N}c_q\sin qz+O(z^{2N})=\chi_N^{(+)}(z)+O(z^{2N}), \quad N\in\mathbb{Z}_{>0}
\end{equation}
and antiperiodic functions
\begin{equation}\label{Approximation_2}
    \frac{1}{z}=\frac{1}{2\sin\frac{z}{2}}-2\sum_{s=\frac{1}{2}}^{N}c^{\prime}_s\sin sz+O(z^{2N+1})=\chi_N^{(-)}(z)+O(z^{2N+1}), \quad N\in\mathbb{Z}_{>0}-\frac{1}{2}
\end{equation}
In both cases it is enough to take $N$ in such a way that $2N\ge p$ for the periodic function and $2N+1\ge p$ for the antiperiodic one, where $p$ is equal to the degree of the maximal pole in the OPE of considered operators:
\begin{equation}
    A(w) B(z)=\frac{\mathcal{O}}{(w-z)^p}+\ldots
\end{equation}
Then we can replace the function $(w-z)^{-1}\rightarrow \chi_N^{(\pm)}(z)$ in the integral (\ref{Lambda}) 
\begin{equation}
    \Lambda_r=\frac{1}{4 i \pi^2} \int_0^{2 \pi} \oint_{\mathcal{C}_z} e^{i r z} \chi_N^{(\pm)}(w-z) \mathcal{T}(A(w) B(z)) d w d z.
\end{equation}
In this formula we take $\chi_N^{(+)}(w-z)$ in two cases: if $A(w)$ and $B(z)$ are both bosonic or if $A(w)$-bosonic and $B(z)$-fermionic; and the function $\chi_N^{(-)}(w-z)$ is taken otherwise. Transforming the integration contour
\begin{equation}
    \int\limits_0^{2 \pi} \oint_{\mathcal{C}_z} d w d z=\mathop{\int\limits_{0}^{2\pi}\int\limits_{0}^{2 \pi}}_{\Im(w)<\Im(z)} d w d z-\mathop{\int\limits_0^{2 \pi}\int\limits_{0}^{2 \pi}}_{\Im(w)>\Im(z)} d w d z,
\end{equation}
we come to the following formula
\begin{equation}\label{Integral}
    \Lambda_r=\frac{1}{4 i \pi^2} \mathop{\int\limits_{0}^{2 \pi} \int\limits_0^{2 \pi}}_{\Im(w)<\Im(z)}\left(e^{i r z} A(w) B(z)+e^{i r w} B(w) A(z)\right) \chi^{(\pm)}_N(w-z) d w d z
\end{equation}
The integral in (\ref{Integral}) can be computed if we move the integration contour over $dw$ to the domain of large negative $\Im(w)$. In this case one can expand
\begin{equation}\label{Chi(+)-expansion}
    -i \chi^{+}_N(w-z)=\frac{1}{2}+\sum_{k=1}^{\infty} e^{i k(z-w)}+2 i \sum_{q=1}^N c_q \sin q(w-z),
\end{equation}
\begin{equation}\label{Chi(-)-expansion}
    -i \chi^{-}_N(w-z)=\sum_{s\in\mathbb{Z}_{>0}-\frac{1}{2}}e^{i s(z-w)}+2 i \sum_{s=\frac{1}{2}}^{N}c^{\prime}_s\sin s(w-z).
\end{equation}
Plugging (\ref{Chi(+)-expansion}), (\ref{Chi(-)-expansion}) into (\ref{Integral}) we come to the following formulae:
\begin{itemize}
    \item $A-$bosonic, $B-$bosonic
    \begin{equation}\label{AB-n-bb}
      (AB)_n=\frac{1}{2}\left(A_0B_n+B_nA_0\right)+\sum_{k>0}
      \left(A_{-k}B_{n+k}+B_{n-k}A_k\right)+\sum_{q>0}c_q
       \left(\big[A_q,B_{n-q}\big]+\big[B_{n+q},A_{-q}\big]\right),
    \end{equation}
    \item $A-$fermionic, $B-$fermionic (here $r\in\mathbb{Z}+\frac{1}{2}$)
     \begin{equation}\label{AB-n-ff}
      (AB)_n=\sum_{r>0}
       \left(A_{-r}B_{n+r}-B_{n-r}A_r\right)+\sum_{s>0}c'_s
       \left(\big\{A_s,B_{n-s}\big\}-\big\{B_{n+s},A_{-s}\big\}\right),
      \end{equation}
    \item $A-$bosonic, $B-$fermionic
         \begin{equation}\label{AB-n-bf}
      (AB)_r=\sum_{k>0}
       \left(A_{-k}B_{r+k}+B_{r-k}A_k\right)+\sum_{q>0}c_q
       \left(\big[A_q,B_{r-q}\big]+\big[B_{r+q},A_{-q}\big]\right),
      \end{equation}
    \item $A-$fermionic, $B-$bosonic (here $s\in\mathbb{Z}+\frac{1}{2}$)
             \begin{equation}\label{AB-n-fb}
      (AB)_r=\sum_{s>0}
       \left(A_{-s}B_{r+s}+B_{r-s}A_s\right)+\sum_{s>0}c^{\prime}_s
       \left(\big[A_s,B_{r-s}\big]+\big[B_{r+s},A_{-r}\big]\right).
      \end{equation}
\end{itemize}
For our purposes it is enough to take the first two terms of the expansions (\ref{Approximation_1}), (\ref{Approximation_2}). In this case the coefficients are 
\begin{equation}
    c_1=-\frac{41}{720}, \quad c_2=\frac{11}{1440}\quad\text{and}\quad
        c^{\prime}_{\frac{1}{2}}=\frac{97}{1920}, \quad  c^{\prime}_{\frac{3}{2}}=-\frac{17}{5760}.
\end{equation}

\bibliographystyle{MyStyle}
\bibliography{MyBib}

\end{document}